\documentstyle[aps,prl]{revtex}

\begin{document}

\draft

\title{Comment on ``Proposal for the Measurement of Bell-Type
Correlations from Continuous Variables''}

\author{K. Banaszek$^{1,2}$,
I. A. Walmsley$^{1}$, and K. W\'{o}dkiewicz$^{2,3}$}

\address{$^1$The Institute of Optics,
University of Rochester,
Rochester, New York 14627\\
$^2$Instytut Fizyki Teoretycznej,
Uniwersytet Warszawski,
PL--00--681 Warszawa, Poland\\
$^3$
Department of Physics and Astronomy
and Center for Advances Studies,
University of New Mexico,
Albuquerque,~New~Mexico 87131}

\date{\today}

\maketitle


\vspace*{5mm}

In a recent Letter \cite{Ralph}, the authors propose a test for
Bell's inequalities based on quadrature measurements for a correlated
parametric source. We present here a local hidden variable (LHV) model
which reproduces all the expectation values predicted in the described
experimental scheme, and point out that the proposal in Ref.~\cite{Ralph}
is based on an inconsistent use of Bell's inequalities. Our model
includes the auxiliary procedure of blocking the squeezed signal input
and measuring the dark noise, which was claimed in Ref.~\cite{Ralph}
to negate the possibility of a local realistic description for the
observed correlations.

Our LHV model is based on the Wigner phase space
representation, which for Gaussian states is well known to provide a local
realistic description of quadrature measurements \cite{Ou}.  We shall
consider six complex hidden variables $E_1,
\ldots, E_6$ described by a joint probability distribution:
\begin{equation}
P( E_1, \ldots , E_6 ) = W_{\text{sq}}(E_1,
E_2) W_{\text{sq}} (E_3, E_4) W_{\text{vac}}
(E_5) W_{\text{vac}} (E_6),
\end{equation}
where 
$W_{\text{sq}}(E,E') = 4 
\exp[-(2+4\chi^2)(|E|^2+|E'|^2)
+ 8\chi\sqrt{1+\chi^2}\text{Re}(EE')]/\pi^2$
is the Wigner function of the two-mode squeezed state,
and $W_{\text{vac}}(E) = 2\exp(-2|E|^2)/\pi$ describes
vacuum fluctuations.
With this choice of hidden variables we can define,
using notation of Ref.~\cite{Ralph}, local realities $X^{\pm}_{k;l}$
for the observables measured by the apparatus $A$ as:
\begin{equation}
X^{j}_{k;l} = 
\left\{
\begin{array}{ll}
e^{i\varphi_l} (E_1 \cos\theta_A + E_3 \sin\theta_A) +
\text{c.c.} \hspace*{5mm}
& \mbox{if $k=A$ and $j=+$} \\
e^{i\varphi_l} (E_3 \cos\theta_A - E_1 \sin\theta_A) +
\text{c.c.}
& \mbox{if $k=A$ and $j=-$} \\
e^{i\varphi_l} E_5 + \text{c.c.}
& \mbox{if $k=va$}
\end{array}
\right.
\end{equation}
where $l=1,2$, $\varphi_1 = 0$, and $\varphi_2 = \pi/2$.  Local
realities for the observables measured by the apparatus $B$ are defined
analogously, with $E_1,E_3$, and $E_5$ replaced respectively by $E_2,
E_4$, and $E_6$. As the Wigner functions $W_{\text{sq}}(E,E')$ and
$W_{\text{vac}}(E)$ are positive definite, this is a perfectly valid LHV
model for the joint detection of quadratures, including
the auxiliary measurement of the dark noise with the blocked squeezed
signal input. A similar hidden variable model can be constructed for
the second scheme discussed in Ref.~\cite{Ralph} involving four squeezed
light sources.

Existence of a local realistic model presented above points out a
conceptual inconsistency of the experimental scheme suggested in
Ref.~\cite{Ralph}.  In fact, the violation of Bell's inequality for
quadrature correlations of a parametric source is a purely artificial
effect generated in the postprocessing of the experimental data. The
Bell inequality used thorough Ref.~\cite{Ralph} was derived originally
by Grangier, Potasek, and Yurke \cite{Grangier} for the measurement
of completely different quantities, namely the intensity correlations
between two light beams measured by photon counting.  Ref.~\cite{Ralph}
instead relies on expressing the joint intensity observables in terms
of quadrature operators. This operation is the critical step which
generates pseudo-nonlocality.  In order to retain the validity of Bell's
inequalities when replacing the intensities with other observables such as
quadratures, one needs to provide a consistent description of both types
of measurements staying exclusively within LHV theories. Specifically, the
local realities $R_A^{i}(\theta_A)$ and $R^j_B(\theta_B)$ representing
count rates should be linked to quadrature realities in a way which
guarantees the positivity of the former, as this is a basic assumption
underlying the Bell's inequality derived in Ref.~\cite{Grangier}.
In contrast, the relation implied by the calculations presented
in Ref.~\cite{Ralph} has the form:
\begin{eqnarray}
R^i_A(\theta_A) & = & (X_{A;1}^i)^2 + (X_{A;2}^{i})^2
- (X_{va; 1}^{i} )^2 - (X_{va ; 2}^{i})^2 \nonumber \\
R^j_B(\theta_B) & = & (X_{B;1}^j)^2 + (X_{B;2}^{j})^2
- (X_{vb; 1}^{j} )^2 - (X_{vb ; 2}^{j})^2.
\end{eqnarray}
When the quadratures appearing on the right hand sides of the above
equations are intepreted as local realities (a legitimate procedure
within LHV theories), there is nothing which guarantees positivity
of $R^i_A(\theta_A)$ and $R^j_B(\theta_B)$. Consequently, the assumptions
underlying the Bell's inequality are invalidated. 

To illustrate the essential difference between measuring
intensities and quadratures, let us note that our Wigner
phase space model which fully explains quadrature correlations in terms
of local hidden variables, fails to provide an analogous description
for the original intensity measurements. The reason for this is that
the Wigner representation of the intensity observables, given by
\begin{equation}
R^{i}_A(\theta_A ) = \left\{
\begin{array}{ll} |E_1 \cos\theta_A + E_3 \sin\theta_A|^2 - {\textstyle
\frac{1}{2}}, \hspace*{5mm} & \mbox{if $i = +$} \\ 
|E_3 \cos\theta_A -
E_1 \sin\theta_A|^2 - {\textstyle \frac{1}{2}}, & \mbox{if $i = -$}
\end{array} \right.  
\end{equation}
for the beam $A$ and similarly for $R^j_B(\theta_B)$, cannot be considered
as local realities because of the negative terms $-\frac{1}{2}$.

The intensity correlations can be related to quadratures only within the
full quantum mechanical description of these quantities.  This creates a
fatal flaw in the proposed scheme: experimental verification of Bell's
inequalities cannot make any use of quantum formalism, as otherwise it
would involve assumptions taken from beyond the range of theories whose
validity it is supposed to test. Clearly, this condition is not satisfied
by the proposal described in Ref.~\cite{Ralph}, which consequently would
not constitute a demonstration of quantum nonlocality.

\end{document}